\documentclass[runningheads]{llncs}
\usepackage{arxiv}

\usepackage{graphicx}
\usepackage{cite}
\usepackage{amsmath,amssymb,amsfonts}
\usepackage{algorithmic}
\usepackage{graphicx,float}
\usepackage{textcomp}
\usepackage{xcolor}
\usepackage{xspace}
\usepackage[labelformat = parens,justification=centering]{subfig}
\usepackage{multirow}
\usepackage{listings}
\usepackage{epstopdf}
\usepackage{hyperref}

\usepackage{cleveref}
\usepackage{todonotes}
\usepackage{url}
\usepackage{svg}

\begin{document}
\title{BDDC preconditioning on GPUs for Cardiac Simulations}
\author{Fritz G\"{o}bel\inst{1} \and
Terry Cojean\inst{1}\orcidID{0000-0002-1560-921X} \and
Hartwig Anzt\inst{2,1}\orcidID{0000-0003-2177-952X}}
\authorrunning{F. G\"{o}bel et al.}
\institute{Karlsruhe Institute of Technology, Germany, \\
           \email{firstname.lastname@kit.edu}
    \and ICL, University of Tennessee, Knoxville, USA}
\maketitle              % typeset the header of the contribution
\begin{abstract}
In order to understand cardiac arrhythmia, computer models for electrophysiology are essential. In the EuroHPC MicroCARD project, we adapt the current models and leverage modern computing resources to model diseased hearts and their microstructure accurately. Towards this objective, we develop a portable, highly efficient, and performing BDDC preconditioner and solver implementation, demonstrating scalability with over 90\% efficiency on up to 100 GPUs.

\keywords{BDDC preconditioning \and Sparse Linear Algebra \and High Performance Computing \and GPUs  \and Cardiac simulations}
\end{abstract}

\section{Introduction}
\label{sec:introduction}

Cardiovascular diseases are the most frequent cause of death worldwide, and half of them are due to cardiac arrhythmia. To understand these disorders of the heart's electrical system very sophisticated and widely used, but currently, they are not powerful enough to take the heart's individual cells into account. Rather than simulating at or below the cell level, the base units of the simulations are groups of hundreds of cells prevents from representing several events in aging and structurally diseased hearts.%, in which reduced electrical coupling leads to large differences in behavior between neighboring cells, with possibly fatal consequences.

Moving towards a cell-by-cell model of the heart increases the size of the problem by 10,000 while making it harder to solve. For this, exascale computers are required, and software leveraging these new architectures, like GPUs, must be developed. The EuroHPC MicroCARD project\cite{potse-pp22,microcard} was started to tackle these challenges. In this paper, we provide efficient, portable, GPU-enabled, and scalable solvers and a BDDC preconditioner tailored for the MicroCARD project. This is implemented within the Ginkgo portable linear algebra framework~\cite{ginkgo-toms} which is the numerical backend selected for the MicroCARD project.

\section{Background and Implementation}
\label{sec:bg_impl}

In the Cell-by-Cell model, individual cells and extracellular space provide a natural division of the simulation domain into subdomains making domain decomposition preconditioners attractive. Considering discontinuous Galerkin discretizations that are required to approximate discontinuous potentials, Huynh et. al. identified Balancing Domain Decomposition by Constraints (BDDC) as a well-suited preconditioner for this type of problem~\cite{cellbycell}.

In a BDDC preconditioner as introduced in~\cite{bddc}, we consider the individual contributions $A_i$ of each subdomain $\Omega_i$ to the global stiffness matrix $A$ locally: $A = \sum_{i = 1}^N R_i^T A_i R_i$ where $R_i$ are adequate restriction matrices. Global coupling is achieved with a coarse problem $A_c = \sum_{i = 1}^{N} R_{ci}^T A_{ci} R_{ci}$ where $A_{ci} = \Phi_i^T A_i \Phi_i$ with $\Phi_i$ arising from the solution of saddle point problems of the form 

\begin{equation}
    \begin{bmatrix} A_i & C_{i}^{T}\\ C_i & 0 \end{bmatrix} \begin{bmatrix} \Phi_i \\ \Lambda_i \end{bmatrix} = \begin{bmatrix} 0 \\ I \end{bmatrix}.
    \label{saddlepoint}
\end{equation}

Here, $C_i$ are constraint matrices on the subdomain boundaries. While different constraint approaches exist, we use the popular simple approach of averaging over subdomain faces and edges while taking full values on corners. Redistributing the solution of the coarse system into the subdomains requires weights for each dof in the subdomains. Here, we use $w_i = \frac{1}{\|\kappa\|}$ where $\kappa$ is the number of subdomains sharing the global dof $i$. In 2D, it will be $1$ for interior nodes, $\frac{1}{2}$ on subdomain edges and $\frac{1}{k}$ on subdomain corners cornering $k$ subdomains.

The preconditioned residual $M^{-1} = v_1 + v_2 + v_3$ has three parts\cite{bddc}: 1) \textbf{coarse grid correction:} as in \cref{coarsegrid} with the coarsened residual $r_c = \sum_{i = 1}^N R_{ci}^T\Phi_i^TW_iR_ir$; 2) \textbf{local subdomain correction:} as in \cref{localsubdomain} with $z_i$ extracted from solving \cref{compzi}; 3) \textbf{static condensation correction:} as in \cref{inner}, where $R_{Ii}$ restricts to the inner dofs of subdomain $\Omega_i$, we compute a new residual $r_1 = r - A(v_1 + v_2)$ with the coarse grid and local subdomain corrections.

    \begin{equation}
        v_1 = \sum_{i = 1}^{N} R_i^T W_i \Phi_i R_{ci} A_c^{-1} r_c
        \label{coarsegrid}
    \end{equation}
    
    \begin{equation}
    v_2 = \sum_{i = 1}^{N} R_i^T W_i z_i
    \label{localsubdomain}
    \end{equation}
    
    \begin{equation}
        \begin{bmatrix} A_i & C_i^T \\ C_i & 0 \end{bmatrix} \begin{bmatrix} z_i \\ \lambda_i \end{bmatrix} = \begin{bmatrix} W_iR_ir \\ 0 \end{bmatrix}
        \label{compzi}
    \end{equation}
    
    \begin{equation}
        v_3 = \sum_{i = 1}^{N}{R_i^TR_{Ii}^T(R_{Ii}A_iR_{Ii}^T)^{-1}R_{Ii}R_ir_1}
        \label{inner}
    \end{equation}

\begin{figure}[!ht]
    \begin{center}
    \subfloat[Weak scaling a 2D-Poisson problem. Local problems: GPU sparse direct solver (solid) or  AMG+GMRES (dashed). \label{fig:weakscaling}]{%
        \includegraphics[width=0.4\textwidth]{"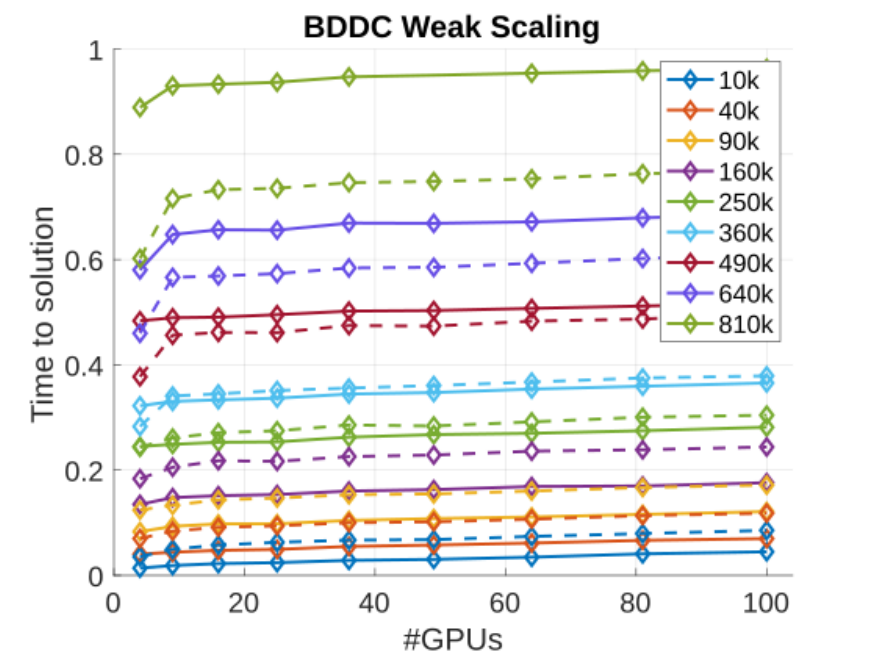"}
    }\hfill
    \subfloat[Weak scaling for BDDC preconditioned vs. plain CG with a local size of 10k dofs per subdomain.\label{fig:bddcvscg}]{%
        \includegraphics[width=0.4\textwidth]{"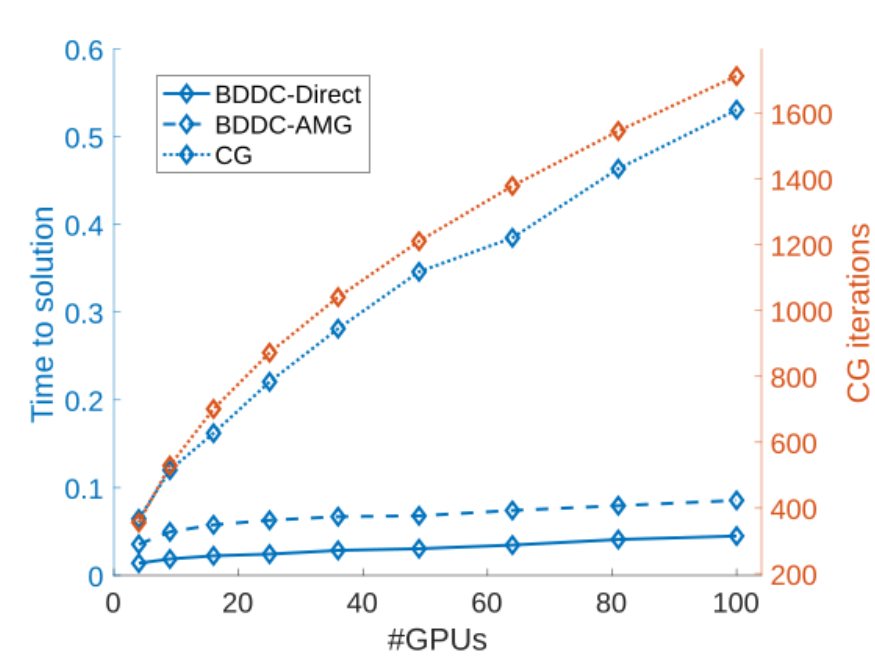"}%
    }
    
    \subfloat[Strong scaling results for a 2D-Poisson stiffness matrix with 6.35 million dofs.\label{fig:strongscaling}]{%
        \includegraphics[width=0.4\textwidth]{"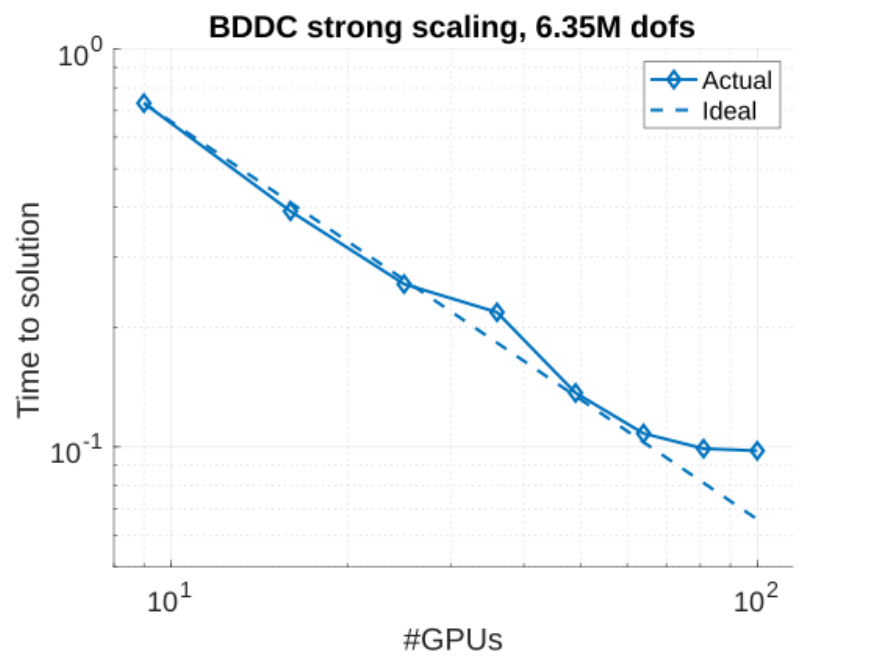"}
    }\hfill
    \subfloat[Strong scaling results for a Bidomain simulation stiffness matrix with 120k dofs.\label{fig:bidomain}]{%
        \includegraphics[width=0.4\textwidth]{"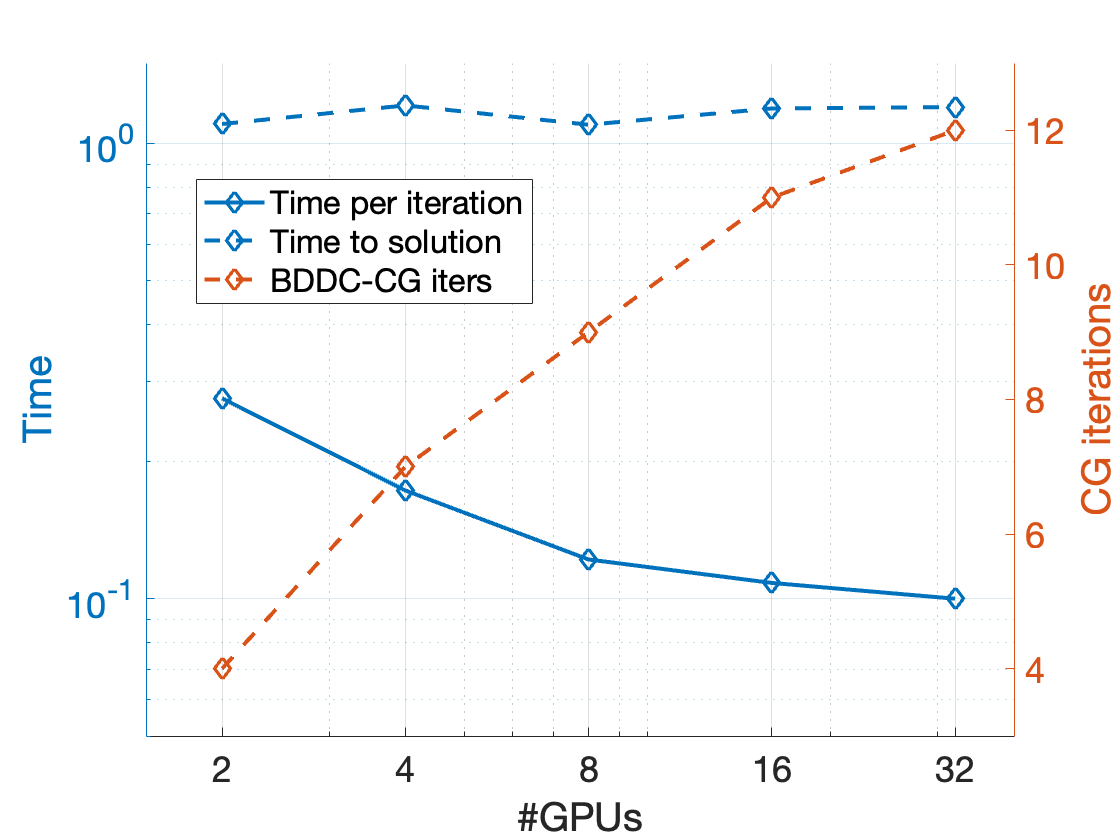"}
    }
    \caption{Results for the BDDC preconditioned CG.}
    \end{center}
\end{figure}

The implementation of the BDDC preconditioner in Ginkgo is individually configurable, the linear solvers used in \cref{saddlepoint,coarsegrid,compzi,inner} can be tuned independently. For the results we show in this paper, we solve the coarse system in (\cref{coarsegrid}) with a plain CG solver and leverage recently developed GPU-resident sparse direct solvers for the solution of the local problems. The direct solvers are used to factorize the complete linear systems in \cref{saddlepoint,coarsegrid,inner} or in the coarsest level of an Algebraic Multigrid (AMG) preconditioner inside a GMRES solver. In both cases, we pre-process the local matrices with MC64~\cite{mc64} for numerical stability and AMD~\cite{amd} for fill-in reduction.

\section{Experimental Results}
\label{sec:experiments}

\Cref{fig:weakscaling,fig:bddcvscg,fig:strongscaling} show scaling results for a 2D-Poisson equation on the unit square, subdivided into equal-sized square subdomains, solved with BDDC preconditioned CG on AMD MI250X GPUs on the Frontier Supercomputer. \Cref{fig:weakscaling} highlights good weak-scaling of our BDDC. When comparing AMG preconditioned GMRES for local problems (dashed lines) against GPU-resident sparse direct solvers (solid lines), we see that AMG is slower for small problems, but gives significant benefits for larger problems. \Cref{fig:bddcvscg} compares CG with a BDDC preconditioner and without using a local size of 10k dofs. Unlike plain CG, we are able to solve the Poisson equation with one global CG iteration when using BDDC leading to almost constant execution time. This confirms that our BDDC implementation is highly effective in improving the convergence and time to solution. \Cref{fig:strongscaling} shows strong-scaling results with 6.35 million dofs. The time to solution scales down until around 9 subdomains in each direction (81 GPUs) where we are no longer able to saturate the GPUs. Finally, \Cref{fig:bidomain} shows strong-scaling results for a matrix obtained from a realistic bidomain simulation of the openCARP~\cite{openCARP}. While the preconditioner deteriorates further from the true inverse of the matrix when increasing the number of subdomains, the timer per iteration scales down for small GPU counts and due to the rather small problem size flattens out earlier than for the larger Poisson example.

\section{Conclusion}
\label{sec:conclusion}

In this paper, we show a portable, high-performance, scalable implementation of the BDDC preconditioner and solvers within Ginkgo~\cite{ginkgo-toms}. This solution can be combined with the ongoing implementation of the new Cell-by-Cell model within OpenCARP~\cite{openCARP} to target Exascale simulations of heart electrophysiology.

\newpage

\section*{Acknowledgement}
This work was supported by the European High-Performance Computing Joint Undertaking EuroHPC under grant agreement No 955495 (MICROCARD). Co-funded by the Horizon 2020 program of the European Union (EU), the French National Research Agency ANR, the German Federal Ministry of Education and Research, the Italian Ministry of economic development, the Swiss State Secretariat for Education, Research and Innovation, the Austrian Research Promotion Agency FFG, and the Research Council of Norway.\\

This is the open access version of the paper published in \cite{europar} as required by the MICROCARD grant agreement.

\newpage
\bibliography{biblio}

\end{document}